\documentclass[10pt,conference,letterpaper,openany]{IEEEtran}
\IEEEoverridecommandlockouts

\usepackage{common/packages}

\IEEEoverridecommandlockouts

\IEEEpubid{\makebox[\columnwidth]{978-3-903176-58-4~\copyright 2023 IFIP \hfill} \hspace{\columnsep}\makebox[\columnwidth]{ }}

\begin{document}

\title{Target Acquired? \\Evaluating Target Generation Algorithms for IPv6}

\author{
	\IEEEauthorblockN{Lion Steger\IEEEauthorrefmark{1}, Liming Kuang\IEEEauthorrefmark{1}, Johannes Zirngibl\IEEEauthorrefmark{1}, Georg Carle\IEEEauthorrefmark{1}, Oliver Gasser\IEEEauthorrefmark{2}}
	\IEEEauthorblockA{\IEEEauthorrefmark{1} Technical University of Munich, Germany}
	\IEEEauthorblockA{\{steger, kuangl, zirngibl, carle\}@net.in.tum.de}
	\IEEEauthorblockA{\IEEEauthorrefmark{2} Max Planck Institute for Informatics, Germany} 
	\IEEEauthorblockA{oliver.gasser@mpi-inf.mpg.de} 
}

\DeclareCiteCommand{\fullcite}
  {\defcounter{maxnames}{99}%
    \usebibmacro{prenote}}
  {\usedriver
     {\DeclareNameAlias{sortname}{default}}
     {\thefield{entrytype}}}
  {\multicitedelim}
  {\usebibmacro{postnote}}

\fancypagestyle{firstpagestyle}{
	\fancyhead[HC]{
		\fbox{
			\begin{minipage}{\linewidth}
				\footnotesize If you cite this paper, please use the TMA reference: \fullcite*{Steger2023Target}
    			\end{minipage}
		}
	}
	\renewcommand{\headrulewidth}{0pt}
}

\maketitle

\thispagestyle{firstpagestyle}

\begin{abstract}
	Internet measurements are a crucial foundation of IPv6-related research.
	Due to the infeasibility of full address space scans for IPv6 however, those measurements rely on collections of reliably responsive, unbiased addresses, as provided \eg by the \hitlist service.
	Although used for various use cases, the hitlist provides an unfiltered list of responsive addresses, the hosts behind which can come from a range of different networks and devices, such as web servers, \ac{cpe} devices, and Internet infrastructure.

    In this paper, we demonstrate the importance of tailoring hitlists in accordance with the research goal in question.
    By using \peeringdb we classify hitlist addresses into six different network categories, uncovering that 42\% of hitlist addresses are in ISP networks.
    Moreover, we show the different behavior of those addresses depending on their respective category, \eg ISP addresses exhibiting a relatively low lifetime.
	Furthermore, we analyze different \acp{tga}, which are used to increase the coverage of IPv6 measurements by generating new responsive targets for scans.
    We evaluate their performance under various conditions and find generated addresses to show vastly differing responsiveness levels for different \acp{tga}.
\end{abstract}

\acresetall

\section{Introduction}
\label{sec:introduction}

The adoption of \ipvsix is continuously increasing, with on average 40\% of all Google users connecting via \ipvsix in March 2023 \cite{google-ipv6}.
Due to the sheer size and sparse population of the \ipvsix address space, exhaustive scans such as in \ipvfour \cite{durumeric2013zmap, adrian2014zippier} are infeasible in the \ipvsix Internet.
Therefore, Internet measurements targeting \ipvsix hosts rely on up-to-date collections of responsive addresses, often known as \textit{Hitlists}.
Moreover, the success of these measurements heavily depends on the quality of their input, reliable targets, and high coverage of the active \ipvsix Internet.
While use cases for such hitlists can vary greatly, hitlists are usually a collection of addresses belonging to different types of devices, such as routers, web servers, or \ac{cpe} devices, treated as a homogeneous set.
This is very inefficient for many measurement studies, as these targets can be expected to be found in completely different network types.
For example, a study on self-hosted video platforms would mainly target educational and company networks, while a study on web content will target vastly different networks, such as \acp{cdn} and hosting providers.
These studies could profit from a categorization of hitlist addresses, as this could allow more focused scans resulting in a reduced scanning overhead and lower load on the network.

The most popular and commonly used \hitlist by \citeauthor*{Gasser2018} \cite{Gasser2018,Zirngibl2022} combines \ipvsix addresses from different sources and performs regular scans to ensure reliable responsiveness.
However, little is known about the current and historic composition of the \hitlist, namely which categories of addresses it contains and whether there is a bias towards \ac{cpe} devices, routers, or servers.
This makes the use of the hitlist unnecessarily difficult and inefficient for many measurement studies.
We address this problem by analyzing the different network categories represented in the data provided by the hitlist service and showing how the categorization of the contained addresses improves the hitlists' usability.

In addition to hitlists, different approaches exist to increase \ipvsix address coverage, \eg by generating new targets.
This is often achieved through so-called \acp{tga}, which employ different methods such as machine learning \cite{Cui2020, Cui2021} and other pattern recognition techniques \cite{Yang2022, Foremski2016}.
Similar to hitlists, little is known about characteristics of \acp{tga} with respect to input from different categories, whether they exbhibit biases towards specific address categories, or whether their results can be improved given more specific input.
Therefore, existing \acp{tga} could benefit from categorizing their input, enabling them to spend their algorithmic and scanning budget on application-tailored target generation.

In this paper, we perform an in-depth analysis of the \hitlist as well as \acp{tga} by categorizing \ipvsix addresses.
This research enables fellow researchers to make better use of the \hitlist and \acp{tga}.
Our contributions in this work are:
\begin{enumerate}
    \item \textbf{\hitlist address categorization:}
        We analyze the \hitlist by \citeauthor*{Gasser2018} with respect to IP address categories. We show that it includes addresses from a variety of categories, \eg \ac{isp} and \ac{nsp} in the input but also the set of responsive addresses, finding a general bias towards \ac{isp} networks with \sperc{42} of responsive addresses.
    \item \textbf{Characterization of address categories:}
        We evaluate whether addresses from differing categories exhibit different behavior over time. We show that addresses from educational and content serving networks are more stable with a median of over 200 days uptime, while \ac{isp} addresses are often only responsive during a single scan. \ac{isp} and \ac{nsp} addresses almost exclusively respond to ICMP, with less than \sperc{10} response rate to any other protocol.
    \item \textbf{Effectiveness analysis of \acp{tga}:}
        We evaluate the effectiveness of different \acp{tga} to identify previously unknown addresses.
        Furthermore, we analyze whether categorized input leads to a change in behavior for \acp{tga}, finding stark contrasts in metrics such as number of generated and responsive addresses and responses to different protocols.
        For example, output generated from the \ac{isp} category has up to \sperc{50} responsiveness, however almost exclusively to ICMP with below \sperc{10} for any other protocol, whereas \ac{cdn} addresses can generate \sperc{65} responsiveness to HTTP.
    \item \textbf{Data and Code:}
        We publish our adaptations to the used \acp{tga}, generated and responsive addresses, analysis scripts and tools used throughout this work, as well as an ongoing categorization of the \hitlist addresses \cite{Steger23TargetData}.
        In order for users of the \hitlist to benefit from our findings, we update the service, include the newly discovered addresses and provide categorized statistics and data to the established service.
\end{enumerate}
\section{Background}
\label{sec:background}

We introduce \acp{tga}, the \hitlist service \cite{Gasser2018} and used data to categorize IP addresses.

\subsection{Target Generation Algorithms}

\begin{table}
	\footnotesize
	\centering
	\caption{List of target generation algorithms with publicly available code used in this work.}
	\begin{tabularx}{0.8\linewidth}{r l l l l}
		\toprule
		Year & Authors & Name & Scanning & Ref \\
		\midrule
		2016 & Foremski et al. & Entropy/IP & Static & \cite{Foremski2016} \\
		2019 & Liu et al. & 6Tree & Dynamic & \cite{Liu2019} \\
		2020 & Song et al. & DET & Dynamic & \cite{Song2022} \\
		2020 & Cui et al. & 6GCVAE & Static & \cite{Cui2020} \\
		2021 & Cui et al. & 6VecLM & Static & \cite{Cui2021a} \\
		2021 & Cui et al. & 6GAN & Static & \cite{Cui2021} \\
		2021 & Hou et al. & 6Hit & Dynamic & \cite{Hou2021} \\
		2022 & Yang et al & 6Graph & Static & \cite{Yang2022} \\
		2022 & Yang et al & 6Forest & Static & \cite{Yang2022a} \\
		2023 & Hou et al. & 6Scan & Dynamic & \cite{Hou2023} \\
		\bottomrule
	\end{tabularx}
	\label{tab:algo_overview}
\end{table}

Discovery of responsive targets for \ipvsix scans is an important task since full address space scans are infeasible.
Besides hitlists, combining targets from existing sources, \eg resolution of domain names, public sources and traceroutes, a variety of so-called \acfp{tga} were developed.
\acp{tga} take a completely different approach to this problem.
They try to identify patterns within existing collections of responsive addresses called the \textit{seed data set}, and generate new targets which are likely to be responsive, called a \textit{candidate set}.
These addresses can be used as input for scans and tested for their responsiveness.
Some of these algorithms also implement their own dynamic scanning mechanisms, which allows them to adapt their search strategy based on intermediate scanning results, and achieve a higher response rate.
\Cref{tab:algo_overview} provides an overview about algorithms we evaluated and used in this work.
These were all the algorithms found in related work which provided publicly accessible source code.

\subsection{IPv6 Hitlist Service}

The \hitlist service from \citeauthor*{Gasser2018} \cite{Gasser2018} was started in 2018 and is maintained since.
It collects IPv6 addresses from different sources and conducts scans for ICMP, TCP/80 (HTTP) and TCP/443 (HTTPS), UDP/53 (DNS) and UDP/443 (QUIC) on a regular basis.
It was updated in 2022 by \citeauthor*{Zirngibl2022} \cite{Zirngibl2022} to improve the quality of the service.
Their hitlist holds over \billion{1.09} unique IPv6 addresses.
Before scanning, they apply different filters, including a blocklist used to ensure opt-out possibilities for networks and ethical scanning, followed by a filter removing addresses which are known to receive bogus DNS injections falsely interpreted as responses.
The injections are linked to the \ac{gfw}, which injects DNS messages regardless if the target host is responsive or not, introducing a strong bias towards DNS responses.
Addresses which do not respond to any other protocol than DNS are therefore filtered.
After this, another \sm{360} addresses are being marked as \emph{aliased} and removed.

\citeauthor*{Gasser2018} \cite{Gasser2018} described aliased prefixes as subnets for which every contained address is mapped to and responded to by one single host, \eg through the \texttt{IP\_FREEBIND} feature of Linux.
\citeauthor*{Zirngibl2022} \cite{Zirngibl2022} showed that some of these prefixes are only fully responsive and used by multiple hosts.
However, each of these prefixes (mostly /64) is infeasible to scan by itself and introduces massive biases of the hitlist.
Therefore, the \hitlist service runs a detection and filters the prefixes.
The list of addresses after all filters is then scanned with probes for different protocols, of which \sm{6.8} were responsive to at least one protocol at March 4, 2023.

\subsection{Categorization}

\peeringdb, run by a community of network-operators, collects information about peerings and interconnections of networks around the world.
Alongside this information, network operators can assign a category to their \ac{as} from twelve categories, including \textit{Content} (Content Delivery Network, short CDN), \textit{Cable/DLS/ISP} (\acl{isp}, short \acs{isp}), \textit{Educational/Research} (Universities, Research Institutes, short EDU), \textit{NSP} (\acl{nsp}, Transit networks) and \textit{Non-Profit} (Non-Profit Organizations, short ORG), which are the five categories relevant in this work.
Alongside \peeringdb, there was an \ac{as} classification by CAIDA, but it was discontinued in 2021 \cite{as_classification}.
Furthermore, \citeauthor*{Ziv2021} \cite{Ziv2021} proposed ASdb in 2021, a system utilizing machine learning approaches to categorize networks at \ac{as} level with high accuracy.
We did however not consider it in our project since their data only goes back to 2021 while the historic data from \citeauthor*{Gasser2018}  \cite{Gasser2018} starts at 2018.

\section{Related Work}
\label{sec:related_work}

The unpredictability of active addresses in the vast IPv6 address space leaves a lot of room for innovative discovery approaches, making the field of \acp{tga} very interesting within \ipvsix research.
Discovery strategies were already described in RFC7707 \cite{RFC7707} based on drafts dating back to 2012.
In 2015, \citeauthor*{Ullrich2015} \cite{Ullrich2015} were among the first to publish on this topic.
They propose an algorithm which iterates through different patterns of a training set, selecting sub-patterns with the highest amount of matching addresses.
New addresses are generated from combining the undetermined bits of the patterns, outperforming the strategies laid out in RFC7707.
In 2016, \citeauthor*{Foremski2016} \cite{Foremski2016} presented Entropy/IP, while \citeauthor*{Murdock2017} \cite{Murdock2017} presented 6Gen in 2017.
The latter identifies dense regions in the input seeds and grows each input address into an independent cluster based on Hamming distance.
All addresses which are inside the clusters and do not belong to the input seeds are regarded as candidate addresses.
The authors claim that 6Gen outperforms Entropy/IP by a factor of 1-8 for identical input data sets.
These early results already show large differences between \acp{tga} and a detailed comparison including different input sets is required.

The remaining \acp{tga} collected for this work (see \Cref{tab:algo_overview}) follow similar approaches.
They extract structural information from IPv6 seed sets and apply different methodologies to improve the quality of generated addresses.
They report different response rates which are hardly comparable.
In 2022, \citeauthor*{Zirngibl2022} \cite{Zirngibl2022} applied four \acp{tga} during their improvement of the \hitlist.
They find that \sixgraph and \sixtree generate the highest number of responsive addresses but do not evaluate algorithms in more detail and different input scenarios.

\citeauthor*{Rye2020} \cite{Rye2020} took a slightly different approach when introducing \emph{edgy} in 2020, focusing on the efficient discovery of the \ipvsix periphery, \ie not servers or clients, but last hop routers.
With \emph{edgy} they were able to discover more than \sm{64} active last hop router addresses.
One year later, \citeauthor*{Li2021} \cite{Li2021} describe a similar approach, discovering more than \sm{50} last hop router addresses through tracerouting non-existent \ipvsix addresses in known or suspected customer subnets of \acp{isp}.
Lastly, \citeauthor*{Beverly2018} \cite{Beverly2018} presented their work focussing on IPv6 topology discovery.
They develop and analyze strategies to collect new interfaces by efficient TTL-limited probing of target address sets, finding \sm{1.3} new router interfaces from their single vantage point.

\section{Data Sources and Target Generation}
\label{sec:methodology}

\begin{figure*}
	\centering
	\footnotesize
	\begin{tikzpicture}[
	entity/.style={draw, minimum width=3cm, minimum height=1cm},
	innerentity/.style={draw, minimum width=2cm,minimum height={1cm}, align=center},
	nameentity/.style={minimum width=2cm},
	node distance=0.4cm
	]%

	\node[nameentity] (scan) {\underline{Scan Results}};
	\node[nameentity, align=center, below = -0pt of scan] (icmp) {ICMP\\TCP/80\\TCP/443\\UDP/443\\UDP/53};

	\node[entity, fit=(scan) (icmp)] (scans) {};
	
	\node[innerentity, right = of scans] (gfw) {GFW \\Filter};
	
	\node[innerentity, left = of scans] (apd) {APD};
	\node[innerentity, left = of apd] (blocklist) {Blocklist};
	
	\node[innerentity, above left = -5pt and 12pt of blocklist] (c1)  {Extracted\\Candidates};
	\node[innerentity, left = of c1] (dynamic)  {Dynamic TGA};
	\node[innerentity, left = of dynamic] (hitlist)  {Full \\Hitlist};
	
	\node[innerentity, below left  = -5pt and 12pt of blocklist] (c2)  {Candidates};
	\node[innerentity, left = of c2] (static)  {Static TGA};
	\node[innerentity, left = of static] (categorized)  {Categorized \\Hitlist};
	
	\draw[-latex, thick] (hitlist) -- (dynamic);
	\draw[-latex, thick] (hitlist) -- (static);
	\draw[-latex, thick] (categorized) -- (dynamic);
	\draw[-latex, thick] (categorized) -- (static);

	\draw[-latex, thick] (dynamic) -- (c1);
	\draw[-latex, thick] (c1) -- (blocklist);
	\draw[-latex, thick] (static) -- (c2);
	\draw[-latex, thick] (c2) -- (blocklist);
	
	\draw[-latex, thick] (blocklist) -- (apd);
	\draw[-latex, thick] (apd) -- (scans);
	\draw[-latex, thick] (scans) -- (gfw);
	
	\draw[-latex,thick] (dynamic) to[in=130,out=50,looseness=2.7] node[midway,below] {Scans} (dynamic);;
	
	\end{tikzpicture}%
	\caption{Pipeline to analyze \acp{tga} (see \Cref{tab:algo_overview}) and their performance within different IP address categories.}
	\label{fig:pipeline}
\end{figure*}
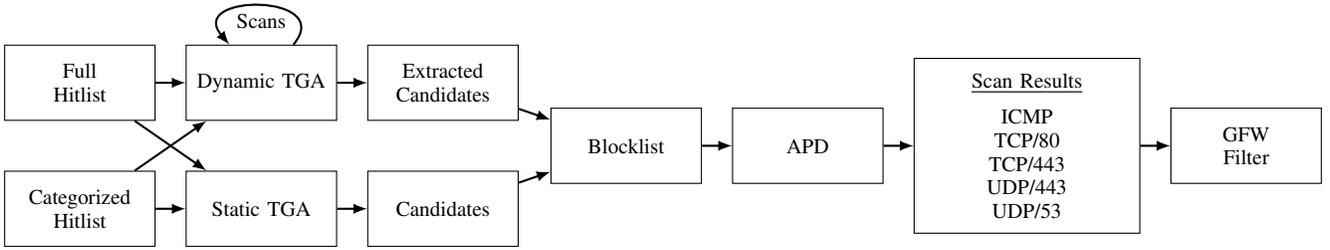

In the following, we describe our data sources and the approach to evaluate the collected \acp{tga} introduced in \Cref{sec:background}.

\subsection{Data Sources}
We use the complete historic data from the \hitlist service \cite{Gasser2018} from July 1, 2018 until March 3, 2023.
We analyze the historic data to gain insights into its categorical composition over time and the responsiveness and stability within each category.
To map addresses to the \ac{as} announcing the respective prefix, we use historic \ac{bgp} Route Views data \cite{routeviews} for one route collector from each scan date.
These mappings to \acp{as} are further used to identify the respective category based on historic \peeringdb data \cite{peeringdb}.

We use the full list of responsive addresses from the \hitlist service from March 3, 2023 as input for \acp{tga} in the following.
Furthermore, we divide the input into filtered versions, \ie the addresses inside the active hitlist filtered based on the \peeringdb categories \emph{Content}, \emph{ISP}, \emph{NSP}, \emph{Non-Profit}, and \emph{Educational}.
While \peeringdb has more network categories, we excluded all categories with less than \sperc{5} representation in the hitlist, additionally including the two categories \emph{Educational} and \emph{Non-Profit} to test the algorithms on smaller seed sets.

\subsection{Target Generation Methodology}
\label{subsec:methodology_tga}

\Cref{fig:pipeline} shows our pipeline to test \acp{tga} on the different input files.
In our study we run and evaluate the ten \acp{tga} listed in \Cref{tab:algo_overview}.
The algorithms run on a machine with an NVIDIA GeForce RTX 2080 GPU, a 24-core Intel Xeon Silver 4214 CPU and \SI{256}{\giga\byte} of RAM.

We run the static \acp{tga} without any modifications apart from input files or output hyper-parameters.
We modify the hyper-parameters number of epochs for \sixgan and the generation budget for \sixgcvae.
To run algorithms in a feasible timeframe, we set the total number of epochs of \sixgan to ten and run \sixveclm with only the first of the predefined temperature hyper-parameters.
\sixgan offers multiple modes for seed classification, of which we choose the \emph{Entropy Clustering} method since the authors report the highest number of generated addresses for this method, which is the metric we optimize for.
Since \sixgan and \sixveclm define the amount of input that is processed via hard-coded values and \entropy is not intended for input greater than \sk{100} addresses, we randomize the input data set with a static, reproducible seed.
Whenever possible we set the output budget to \sm{10}, since we wanted to keep the size of the candidate sets at a similar scale to the input, \ie the hitlist.
We implement the approach described by the authors of \sixgraph to generate candidates from the dense regions identified by their algorithm.
For \sixforest, this process is not fully described or implemented.
Therefore, we follow the same procedure as for \sixgraph, generating distance-based targets and additionally generating full combinations if the number of wildcards, \ie free dimensions, is smaller than four.

Before running the algorithms with dynamic scanning capabilities, we modify them \first to conform to our scanning parameters and \second to output not only the addresses responsive to their scans but also the addresses which they probed, \ie the addresses which they consider to be in the \emph{candidate set}.
We do this because the integrated scanning mechanisms of the algorithms only scan with ICMP probes, while we want to apply our own scanning mechanisms with a variety of protocol probes.
Furthermore, in order to compare the response rate of both dynamic and static algorithms, we have to scan the candidate sets for both instead of only the results of the dynamic algorithms.
In order to achieve consistency with the static algorithms, we run each dynamic \acs{tga} with the same \sm{10} scanning budget.

After the successful generation of all candidate sets, we combine them into one target file.
As a first step, the target file is stripped of duplicates and filtered by applying a blocklist, which we actively maintain in order to adhere to ethical scanning guidelines (see \Cref{sec:ethics}).
Next, we conduct aliased prefix detection, as described by \citeauthor*{Gasser2018} \cite{Gasser2018} and additionally use the known aliased prefixes from the \hitlist service as a filter.
The remaining, non-aliased addresses are then used as input to the \zmap{}v6 port scanning tool\footnote{\url{https://github.com/tumi8/zmap}}.
We scan from a single vantage point in Munich, Germany, connected to the German National Research and Education Network.
As a last step, the responses to the UDP/53 (DNS) probes are passed through a GFW filter, which we describe in \Cref{subsec:evaluation_gfw}.

During the analysis stage, we take the following steps:
First, all duplicates and overlaps with the respective seed set are removed for all candidate sets.
Then the filtered candidate sets and filtered scan file are matched to identify the responsive portions of the candidate sets.
This provides the final result for each \acs{tga}.

\subsection{Ethical Considerations}
\label{sec:ethics}
During this work, we strictly follow ethical considerations for scanning as described in \cite{menloreport, PA16}.
We limit the rate of all scans, apply a blocklist and filter aliased prefixes based on our own detection but also the list of published aliased prefixes by \citeauthor*{Gasser2018} \cite{Gasser2018}.
We evaluated dynamic \acp{tga} whether they adhere to our scan limits and executed them in an environment where we can monitor their behavior and apply our own blocklist.
We inform about our scans based on reverse DNS, a website hosted on the scanning machine and in WHOIS.
We respond to all opt-out requests and add address ranges to our internal blocklist.

\section{Results}
\label{sec:results}
We analyze the \hitlist composition with regard to different network categories, compare different \acp{tga}, and investigate the influence of the \acs{gfw} on our measurements.

\subsection{Hitlist Categorization}
\label{subsec:results_historical}

\begin{figure}
	\centering
	\includegraphics{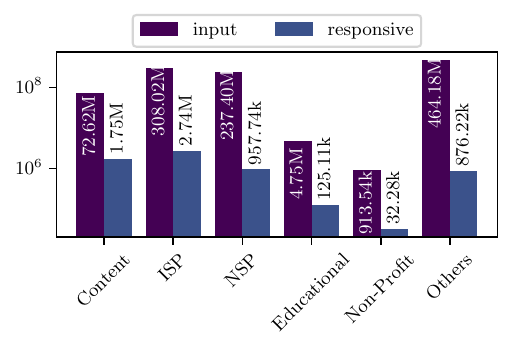}
	\caption{Prevalence of different categories in the \hitlist on March 3, 2023. Note the logarithmic y-axis.}
	\label{fig:hitlist_cats}
\end{figure}

\begin{figure}
	\centering
	\includegraphics{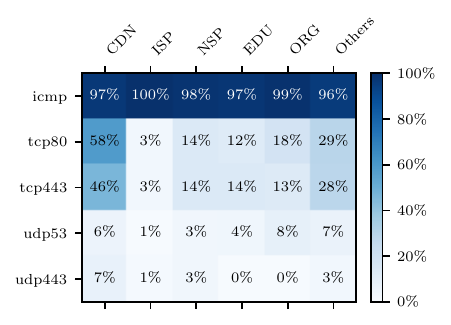}
	\caption{Responses to the different protocols per category within the \hitlist on March 3, 2023.}
	\label{fig:heatmap_proto_cat}
\end{figure}

First, we analyze the different network categories of the \hitlist's input and responsive addresses.
The network categories represented in the \hitlist show different prevalence and behavior.
\Cref{fig:hitlist_cats} shows the distribution of addresses across categories in the full hitlist input as well as its responsive part. The responsive addresses as well as the full hitlist are dominated by \ac{isp} and \ac{cdn} addresses, with almost \sperc{50} combined.

Next, we analyze the responsiveness in more detail, by looking at different probe protocols and network categories.
\Cref{fig:heatmap_proto_cat} shows how many protocol-specific responses the latest scan receives per category, relative to the total number of IP addresses per category which responded to at least one protocol probe.
Addresses belonging to \acp{cdn} have the highest relative number of responses to HTTP and HTTPS probes, with a low, but still comparatively large number of QUIC responses.
This is expected, since web hosting via HTTP/S is one of the primary functions of \acp{cdn}, which are also among the first to deploy QUIC at scale \cite{Zirngibl2021}.
\ac{isp} addresses, on the other hand, show almost no response to any protocol other than ICMP.

The high total share of \ac{isp} addresses in the hitlist, together with the low response rate to any protocol other than ICMP shows the importance of categorizing hitlists before using them as input for application specific scans, as a large part of scanning traffic can be avoided by carefully selecting target addresses from specific categories.

\begin{figure}
	\centering
	\includegraphics{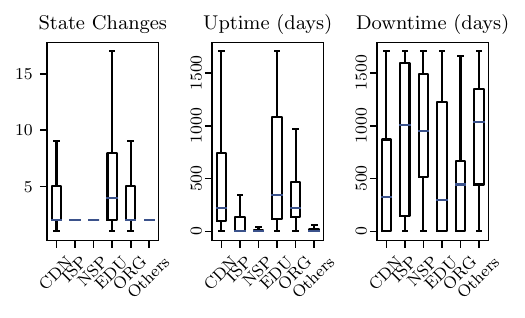}
	\caption{Stability of responsive IPv6 addresses of the \hitlist per category. Data from scans is used from July 2018 to March 2023, while addresses newly discovered from November 2022 onwards are excluded to reduce the impact of newer scans.}
	\label{fig:ipstability}
\end{figure}

To better understand the stability of addresses within the \hitlist, we analyze the categories represented in the hitlist over time using three \emph{IP stability} metrics.
First, the number of \emph{state changes}, \ie the number of times an IP was added to or removed from the responsive part of the hitlist.
This can be seen as a lower-bound for the times an IP address changes between online and offline.
Second and third, we look at the summed up number of uptimes and downtimes of each address, starting when an IP address is first added to the hitlist, and ending at the time of analysis.
These three metrics combined make up the \emph{IP stability} of an address over time.
A stable IP address has a small number of state changes, high uptime, and low downtime, the opposite being true for an unstable IP address.
For this analysis, in order to avoid analyzing IP addresses with not enough historic data, we exclude all addresses added to the hitlist within the last 100 days.

\Cref{fig:ipstability} shows that the different network categories exhibit very distinctive behavior in IP stability.
Most addresses from the categories \ac{isp}, \ac{nsp}, and Others have exactly one state change, but median uptimes of less than a week, meaning that they are included once in the responsive hitlist for seven days and never again afterwards.
ISP networks often use prefix rotation to avoid tracking of their clients and enhance their privacy \cite{Rye2021,saidi2022one}, which means that devices like home routers often change IP address.
Including them in hitlists leads to an increase in unstable targets, which is underlined in the results of this analysis.
This also applies to NSP networks, which offer similar services and partly to IP addresses in the Others category.
In contrast to this, addresses in CDN, Educational, and Non-Profit networks have much higher uptimes, as addresses hosting content have to be available reliably.
The higher number of state changes in these networks can be due to maintenance periods or changes in ownership of the respective servers.

This means that for longitudinal measurement studies which focus on protocols other than ICMP, addresses from categories such as \ac{nsp} and \ac{isp} should be used with care, as they have only very limited periods of responsiveness and generally respond less to protocols other than ICMP, compared to addresses from Content Delivery, Educational or Non-Profit Organization networks.

\subsection{Target Generation Algorithms}

        \begin{table*}
            \centering
            \setlength{\tabcolsep}{3pt}
            \caption{Amount of candidate (cand.) and responsive (resp.) addresses generated by the algorithms when using different categories as well as the full hitlist as seed data set.}
            \label{tab:algo_rates_details}
            \begin{tabularx}{0.97\linewidth}{ X r r r r r r r r r r r r r r r r r r r r }
                \toprule
                  & \multicolumn{2}{c}{6Forest} & \multicolumn{2}{c}{6GAN} & \multicolumn{2}{c}{6GCVAE} & \multicolumn{2}{c}{6Graph} & \multicolumn{2}{c}{6Hit} & \multicolumn{2}{c}{6Scan} & \multicolumn{2}{c}{6Tree} & \multicolumn{2}{c}{6VecLM} & \multicolumn{2}{c}{DET} & \multicolumn{2}{c}{Entropy} \\
\cmidrule(lr){2-3} \cmidrule(lr){4-5} \cmidrule(lr){6-7} \cmidrule(lr){8-9} \cmidrule(lr){10-11} \cmidrule(lr){12-13} \cmidrule(lr){14-15} \cmidrule(lr){16-17} \cmidrule(lr){18-19} \cmidrule(lr){20-21}  & cand. & resp. & cand. & resp. & cand. & resp. & cand. & resp. & cand. & resp. & cand. & resp. & cand. & resp. & cand. & resp. & cand. & resp. & cand. & resp. \\
                \midrule
                Content & 2M & 174k & 487k & 13k & 3M & 14k & 35M & 443k & 10M & 231k & 9M & 491k & 11M & 417k & 78k & 4k & 9M & 361k & 6M & 8k \\
                ISP & 3M & 2M & 410k & 55k & 845k & 179k & 25M & 3M & 8M & 3M & 8M & 4M & 11M & 3M & 18k & 2k & 8M & 3M & 6M & 1M \\
                NSP & 2M & 128k & 521k & 4k & 3M & 15k & 31M & 527k & 10M & 552k & 9M & 884k & 9M & 1M & 66k & 6k & 2M & 382k & 6M & 16k \\
                Educational & 1M & 19k & 316k & 3k & 700k & 585 & 2M & 22k & 24M & 100k & 10M & 38k & 11M & 107k & 84k & 1k & 1M & 745 & 4M & 3k \\
                Non-Profit & 711k & 39k & 125k & 9k & 284k & 3k & 296k & 15k & 20M & 3M & 10M & 946k & 8M & 2M & 0 & 0 & 6M & 356k & 4M & 14k \\
                Full & 2M & 494k & 486k & 41k & 2M & 111k & 106M & 5M & 18M & 3M & 6M & 2M & 35M & 5M & 49k & 4k & 8M & 1M & 6M & 59k \\
                \bottomrule
            \end{tabularx}
        \end{table*}

\begin{figure}
	\centering
	\includegraphics{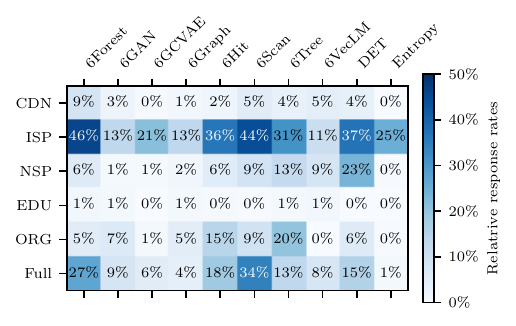}
	\caption{Relative response rate for every candidate set generated by the \acp{tga} with different categorized input sets as well as the full hitlist as input.}
	\label{fig:result_hitrates_full}
\end{figure}

\begin{figure}
	\centering
	\includegraphics{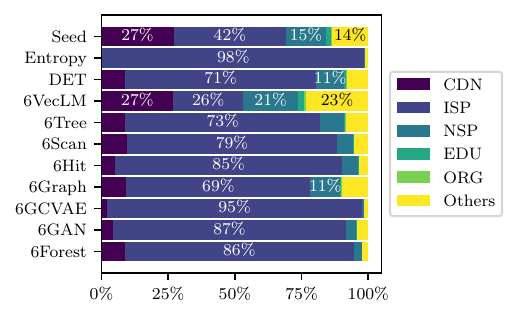}
	\caption{Category distribution of responsive addresses from each \acl{tga} applied on the full set of responsive addresses from the \hitlist.}
	\label{fig:result_cat_distr_full}
\end{figure}

Given the diverse composition of the \hitlist with respect to covered network categories and the differences in responsiveness and IP stability for these categories, we set out to analyze the properties of \acp{tga}.
We evaluate the generated candidate sets of all \acp{tga} in two \zmap scans on March 17 and March 23, 2023 as described in \Cref{subsec:methodology_tga}.
The scans combined the generation based on the full set of responsive addresses as well as the categorized input sets.
The second scan was conducted due to an error in the first one, through which the candidate sets of the dynamic algorithms were not included.
For this second scan, including only the candidate sets from the dynamic altorithms, \ac{apd} was replaced with a filter for known aliased prefixes from the \hitlist service.
We merge the results from both scans and present the results in the following, together with the different metrics used for the evaluation.

\niceparagraph{Generation rate and candidate set size}
The various algorithms generate vastly different numbers of candidate addresses.
Moreover, the generated addresses are also highly dependent on the used seed set.
As already elaborated, the prevalence of categories in the hitlist and therefore the size of the categorized seed sets vary (see \Cref{fig:hitlist_cats}).
Therefore, we compare the \emph{generation rate} of the algorithms, \ie the size of the candidate set relative to the seed set size.
Algorithms like \sixveclm and \sixgan have relatively low generation rates, which is due to the fact that these algorithms limit the amount of processed input to a hard coded value (see \Cref{sec:methodology}).
Algorithms such as \sixgraph can generate more than \sm{100} candidate addresses, which is \sperc{1638} of the size of the seed sets.
With the Non-Profit seed set, \sixveclm is unable to generate a candidate set, as the seed set is smaller than the predefined input size, which we could not successfully modify.
For the exact sizes and generation rates, see Appendix \Cref{tab:algo_rates_overview}.

If the scanning budget of a measurement is critical, large candidate sets should either be sampled or another algorithm should be chosen. If large candidate sizes are desired, large inputs or algorithms with high generation rates are best suited.

\niceparagraph{Response rates}
Internet measurement studies are not only dependent on a scanning budget, but also strive to avoid unnecessary probes which are unlikely to trigger responses.
Therefore, it is important to analyze the response rate, \ie the portion of addresses which responds to at least one protocol, for the different candidate sets.
As can be seen in \Cref{fig:result_hitrates_full}, a larger candidate set does not lead to a higher response rate.
Instead response rates are more strongly linked with the input set category as well as the difference between dynamic and static algorithms.
Dynamic algorithms, due to their ability to adapt their generation strategy based on the results of their scans, have among the top response rates for all categories, up to \sperc{45} for some.
On the other hand, static algorithms rarely show response rates over \sperc{15}, with \sixforest being one of the few exceptions.
Using ISP addresses as input shows the best response rates for almost all algorithms, even better than with uncategorized input.
Candidate addresses generated from educational networks, on the contrary, have the lowest response rate at hardly over \sperc{1} for any algorithm.

Again, for measurements with limited scanning budget, choice of input is shown to be critical for the efficiency of the \acp{tga} and subsequently the scans.

\niceparagraph{Category distribution in responsive addresses}
While all \acp{tga} receive the same input, not only do their candidate sets vary greatly in size, but also in the distribution of represented network categories.
\Cref{fig:result_cat_distr_full} shows the category distributions in the candidate sets generated by the algorithms and the seed set when using the full hitlist as input.
Most algorithms show a strong bias towards ISP addresses, which are also present in the seed data set, although at a much lower percentage.
Especially the relatively small percentage of generated CDN addresses is in stark contrast to the ratios of the seed set.
When using categorized input, all but two algorithms generate \num{95}--\sperc{100} of their addresses in the same category as the input.
The only exceptions are \sixgcvae and \entropy, which generate up to \sperc{62} and \sperc{13} from other categories for some inputs, respectively.
These findings show the need to filter the input to \acp{tga} depending on the desired use case, as the algorithms exhibit a strong bias towards some categories.

\begin{figure}
	\centering
	\includegraphics{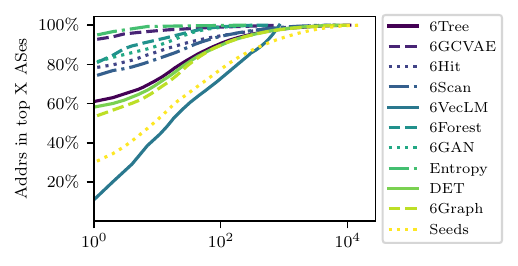}
	\caption{Cumulative AS distribution of the responsive candidate sets generated by the algorithms using the full hitlist as input. Note the logarithmic x-axis.}
	\label{fig:as_results_full}
\end{figure}

\niceparagraph{AS origin distributions}
While categorization on an \acs{as} level via \peeringdb already gives us some information of the origin of the contained addresses, the exact AS distributions still hold some more information.
The cumulative AS distribution of the candidate sets generated from the full hitlist are shown in \Cref{fig:as_results_full}.
Most candidate sets generated by the \acp{tga} are more biased towards single \acp{as}:
The majority of \acp{tga} contain \num{50}--\sperc{95} addresses from a single AS, whereas the top ten ASes of the seed set cover only around \sperc{50} of their addresses.
The most popular AS for all but one candidate sets is AS12322 (FREE SAS, an \ac{isp} network from France).
The only exception to this bias is the candidate set of \sixveclm, which contains only five addresses from AS12322 and is even more evenly distributed among ASes than the seed set dataset.
While AS12322 is also the AS with the highest share in the seed data set, it covers only \sperc{30} of it.
Looking at the structure of the addresses from this AS responding to ICMP, it is visible that over \sperc{99} of them have the host part set to \texttt{::1}.
They are all within the same /39 prefix and only differ in the 10 to 15 nibble of the address.
This a very clear structure which has easy to detect patterns, ideal for discovery by \acp{tga}.
Addresses from this AS were first added to the hitlist via CT logs and the Bitnodes dataset \cite{Gasser2018} and their share drastically increased with the first usage of \acp{tga} in the 2022 paper \cite{Zirngibl2022}.
The high percentage of addresses from AS12322 in most candidate sets also explains their bias towards the ISP network category.
This further stresses the need to filter certain categories for use cases where addresses from the respective categories should not be targeted.
Furthermore, also addresses from specific ASes should be filtered, as their inclusion in the seed set can introduce biases towards those networks far beyond their presence in the seed set.

\niceparagraph{Number of covered ASes}
Next to the distribution of the ASes contained in the candidate sets, their absolute number is equally relevant.
\acp{tga} should generate new targets which represent the active part of the IPv6 Internet, which cannot be achieved if only a small number of ASes is covered in their candidate sets.
Most candidate sets cover substantially fewer total ASes than the respective seed sets, especially when using the full hitlist as input.
Only in very specific circumstances, when the seed set already contains very few ASes (such as for Educational or NSP), some candidate sets cover more ASes.
Very low coverage rates compared to the seed set means that algorithms discover ASes from very specific origins which cannot represent the \ipvsix Internet.
Even when combined, the candidate sets of all algorithms only cover \sperc{75} of the ASes in the seed set.
While the combined candidate sets include 684 ASes which have not been covered by the seed set, \num{4875} ASes from the seed set are not included.
The exact number of newly covered ASes can be seen in \Cref{tab:algo_rates_overview}.

It is therefore imperative to use additional address sources in order to achieve a higher AS coverage if measurements should cover a representative subset of the \ipvsix Internet.
It also raises the questions if current \acp{tga} adequately address the need for a balanced candidate set across a large number of ASes.

\niceparagraph{Ratio of aliased prefix}
Aliased prefixes, as defined in \Cref{sec:background}, are excluded in our scans as they do not add any valuable information, but instead introduce a bias to the results.
It is therefore an important measure of quality for a candidate set to contain few addresses from aliased prefixes.
As described in \Cref{subsec:methodology_tga}, we conducted APD ourselves and with the aliased prefixes published by the \hitlist service.
We compared the unfiltered sets with non-aliased versions and found that most algorithms have a negligible rate of addresses from aliased prefixes, thereby not impacting the algorithms candidate set quality when filtered.
Two exceptions are \sixgcvae and \entropy, which generate up to almost \sperc{50} aliased addresses for some categorized input as well as the full input.
This decreases the usable size of their candidate sets substantially, which should be kept in mind before scanning.
The exact rate of aliased prefixes can be found in Appendix \Cref{tab:algo_rates_overview}.

Although the rate of aliased prefixes in most candidate sets were relatively low, which means that only little unnecessary scanning overhead would be introduced, we still stress the need for \ac{apd}.

\begin{figure}
	\centering
	\includegraphics{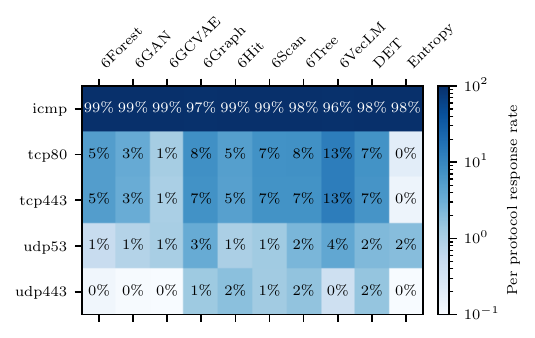}
	\caption{Response rates to the different protocols per algorithm generated on full hitlist input. Note the color map log scale.}
	\label{fig:algo_proto_rate}
\end{figure}

\niceparagraph{Protocol responses}
Depending on the use case for the generated addresses, it can be crucial to discover targets with a high response rate to a certain protocol.
\Cref{fig:algo_proto_rate} shows the response rate to the different protocols per candidate set.
All candidate sets have the highest response rate to ICMP probes, which is to be expected due to the prevalence in the seed set.
Moreover, unlike in IPv4, ICMP in IPv6 can not simply be fully blocked due to its important functionality in stateless address autoconfiguration \cite{rfc4862}.
Responses to other protocols are much less frequent for all candidate sets.
Especially the response rate for HTTP and HTTPS is very similar to the share of non-ISP addresses in the responsive portion of the candidate sets.
Looking at \Cref{fig:result_cat_distr_full}, we can see that the responses to the candidate set of \sixveclm have the lowest share of ISP addresses and the highest number of responses to HTTP and HTTPS.
\entropy and \sixgcvae on the other hand, have more than \sperc{95} ISP addresses in their respective responses and the lowest share of protocol responses other than ICMP.
The per-protocol response rate for the candidate sets generated with categorized inputs show a very strong correlation with the per-category response rates of the hitlist, see \Cref{fig:heatmap_proto_cat}.
With CDN input addresses, all candidate sets receive between 30 and \sperc{65} HTTP and HTTPS responses, whereas with ISP input, no candidate set generates more than \sperc{3} response rate for any protocol besides ICMP.

This shows that input and algorithm should be chosen carefully depending on the desired protocol responses for the use case.

\subsection{GFW Filtering}
\label{subsec:evaluation_gfw}

As described in \Cref{fig:pipeline}, the responses to our DNS probes are post-processed with a \emph{GFW Filter}.
Our probes contain \texttt{AAAA} queries for \texttt{www.google.com} and frequently receive responses with addresses from the \emph{Teredo} prefix in their answer section.
These responses do not originate from legitimate hosts, but are instead likely injected by the \ac{gfw} for reasons of censorship \cite{Zirngibl2022}.
\texttt{google.com} is on the list of censored domains in China \cite{hoang2021gfwatch} and no web service for \texttt{www.google.com} is reachable at the returned addresses.
Following the description of \citeauthor*{Zirngibl2022} \cite{Zirngibl2022}, we filter all responses containing addresses from the \emph{Teredo} prefix in their answer section and do not count them as responsive in our work.

In both our scans, however, we see a substantial change in the format of the injections, as we receive responses containing addresses from Facebook's network in their answer section.
This change in behavior indicates that the \ac{gfw} tries to adapt the \ipvsix injections to the format of their \ipvfour injections, which contain addresses from a fixed pool of \ipvfour addresses, including addresses from Facebook and Twitter \cite{anonymous2020tripletcensors}.
Until recently, every returned Teredo address encoded a corresponding address from the \ipvfour pool in the last 32 bits of the address, whereas now, a separate pool of \ipvsix addresses from similar networks such as Facebook is being returned.
We argue that, at this moment and for probes querying \texttt{www.google.com}, filtering out responses containing addresses from Facebook is sufficient, as our scans show similar response rates to DNS probes as the \hitlist service.
To the best of our knowledge, this new behavior of the \ac{gfw} has not been documented before.
DNS scans targeting \ipvsix addresses need to take this behavior into account and adjust the filtering pipeline accordingly.
Since we, however, expect that the \ac{gfw} can change its behavior in unpredictable ways, we chose not to adapt the filter of the \hitlist service to this new type of injection, and instead changed the domain name which is used in the regular query probes.
The new domain name is not censored by the \ac{gfw} and does not trigger any injections, which we expect to remain the same in the future.
We argue that this yields more stable and usable results for researchers using the \hitlist.

\section{Discussion and Conclusion}
\label{sec:conclusion}

In this work, we have highlighted the dependency of IPv6 measurements on their targets.
We have shown that address collections such as the \hitlist service contain multiple types of networks, including \acp{isp}, \acp{cdn}, and \acp{nsp}, with different behavior.
While addresses from \ac{cdn} networks respond to HTTP and HTTPS at a rate of around \sperc{50} and are responsive for a median time span of more than 200 days, ISP addresses are mostly only available for a single scan and only respond to ICMP for \sperc{97} of the addresses.
Furthermore, we evaluated the behavior of different \aclp{tga}, using the full and categorized versions of the hitlist as input.
We demonstrated that the input has a strong influence on various metrics, such as the number of generated and responsive addresses, protocol responses, and  addresses origin.
All but one candidate sets generated from uncategorized input show a very strong bias towards ISP networks, which in turn have a strong bias towards single \acp{as} and generally have a response rate below \sperc{10} for any protocol other than ICMP.
Output from categorized seed sets consists of addresses from the respective input category, exhibiting behavior similar to the addresses from the respective categories in the hitlist.
However, we learned that most \acp{tga} are complex tools and the majority of published tool chains are trained and optimized on a specific input.
While we tried to adapt the algorithms and parameters to suite our use cases and scenarios, we were not able to reach published rates of responsive addresses.
Furthermore, algorithms with dynamic scanning capabilities are not suited for all use cases, as the adherence to scanning rates, blocklists and detection of aliased prefixes cannot be achieved without modifications.
Our work provides a detailed comparison under different circumstances to allow for a selection of suitable \acp{tga} and a more focused analysis and optimization in the future.
As an example, a scan application which requires large numbers of targets and does not have tight restraints on scan budgets, should opt for an algorithm such as \sixgraph or \sixtree, as they generate the largest candidate sets.
Scenarios, on the other hand, which dictate efficient scanning with a high response rate and do not require modifications to the candidate set before scanning, are best suited for dynamic algorithms, as they reach the highest response rates.

Future IPv6 Internet measurements are encouraged to use our findings to increase the efficiency of their scans by removing unnecessary scanning overhead and generating targets better suited for their use case.
Researchers conducting IPv6 measurements should keep in mind that the current hitlist shows a bias towards \ac{isp} addresses.
These addresses are only short lived and should therefore not be used for long-term studies.
A proper selection of scan specific targets from the hitlist and a proper application of \acp{tga} on specific seed sets can however improve future scans and reduce unnecessary probing.

\section*{Acknowledgments}

This work was partially funded by the German Federal Ministry of Education and Research under the projects PRIMEnet (16KIS1370), 6G-life (16KISK002) and 6G-ANNA (16KISK107). 

\renewcommand*{\bibfont}{\normalfont\footnotesize}
\printbibliography

        \begin{table*}[b]
            \centering
            \caption{Appendix: Overview of different metrics for the algorithm per categorized input.}
            \label{tab:algo_rates_overview}
            \begin{tabularx}{1.01\linewidth}{ l l r r r r r r r r r r }
                \toprule
                 &  & 6Forest & 6GAN & 6GCVAE & 6Graph & 6Hit & 6Scan & 6Tree & 6VecLM & DET & Entropy \\
                \midrule
                \multirow{6}{5em}{Number of candidate addresses} & Content & \sm{1.94} & \sk{486.59} & \sm{3.28} & \sm{34.85} & \sm{10.02} & \sm{8.94} & \sm{11.08} & \sk{77.82} & \sm{8.56} & \sm{5.87} \\
                 & NSP & \sm{2.09} & \sk{521.17} & \sm{2.60} & \sm{30.95} & \sm{9.71} & \sm{9.48} & \sm{9.16} & \sk{66.36} & \sm{1.63} & \sm{5.52} \\
                 & Educational & \sm{1.44} & \sk{316.24} & \sk{699.80} & \sm{1.98} & \sm{24.02} & \sm{9.88} & \sm{11.05} & \sk{83.54} & \sm{1.06} & \sm{4.48} \\
                 & Non-Profit & \sk{710.61} & \sk{125.30} & \sk{284.16} & \sk{295.57} & \sm{19.57} & \sm{9.97} & \sm{7.89} & \num{0} & \sm{5.87} & \sm{3.79} \\
                 & ISP & \sm{3.33} & \sk{409.52} & \sk{845.24} & \sm{24.57} & \sm{8.37} & \sm{7.91} & \sm{11.17} & \sk{18.24} & \sm{7.79} & \sm{5.98} \\
                 & Full & \sm{1.84} & \sk{486.22} & \sm{2.00} & \sm{106.12} & \sm{17.92} & \sm{5.87} & \sm{35.39} & \sk{48.55} & \sm{8.30} & \sm{5.63} \\
                \midrule
                \multirow{6}{5em}{Generation factor} & Content & \num{1.11} & \num{0.28} & \num{1.88} & \num{19.97} & \num{5.74} & \num{5.13} & \num{6.35} & \num{0.04} & \num{4.90} & \num{3.36} \\
                 & NSP & \num{2.18} & \num{0.54} & \num{2.72} & \num{32.29} & \num{10.13} & \num{9.89} & \num{9.55} & \num{0.07} & \num{1.70} & \num{5.76} \\
                 & Educational & \num{11.48} & \num{2.53} & \num{5.60} & \num{15.82} & \num{192.22} & \num{79.09} & \num{88.38} & \num{0.67} & \num{8.48} & \num{35.83} \\
                 & Non-Profit & \num{22.01} & \num{3.88} & \num{8.80} & \num{9.15} & \num{605.98} & \num{308.69} & \num{244.39} & \num{0.00} & \num{181.82} & \num{117.42} \\
                 & ISP & \num{1.21} & \num{0.15} & \num{0.31} & \num{8.91} & \num{3.03} & \num{2.87} & \num{4.05} & \num{0.01} & \num{2.82} & \num{2.17} \\
                 & Full & \num{0.28} & \num{0.08} & \num{0.31} & \num{16.38} & \num{2.77} & \num{0.91} & \num{5.46} & \num{0.01} & \num{1.28} & \num{0.87} \\
                \midrule
                \multirow{6}{5em}{Number of responsive addresses} & Content & \sk{173.90} & \sk{13.42} & \sk{13.97} & \sk{443.44} & \sk{230.62} & \sk{491.22} & \sk{416.75} & \sk{3.87} & \sk{360.54} & \sk{7.62} \\
                 & NSP & \sk{128.27} & \sk{3.66} & \sk{15.31} & \sk{527.19} & \sk{552.35} & \sk{884.09} & \sm{1.15} & \sk{5.68} & \sk{381.85} & \sk{15.86} \\
                 & Educational & \sk{19.37} & \sk{2.62} & \num{585} & \sk{22.04} & \sk{99.82} & \sk{37.73} & \sk{106.52} & \sk{1.23} & \num{745} & \sk{2.59} \\
                 & Non-Profit & \sk{38.91} & \sk{8.50} & \sk{2.61} & \sk{15.12} & \sm{2.86} & \sk{946.16} & \sm{1.58} & \num{0} & \sk{355.65} & \sk{13.66} \\
                 & ISP & \sm{1.53} & \sk{55.22} & \sk{179.00} & \sm{3.27} & \sm{3.03} & \sm{3.50} & \sm{3.45} & \sk{2.06} & \sm{2.89} & \sm{1.49} \\
                 & Full & \sk{494.21} & \sk{41.36} & \sk{111.48} & \sm{4.74} & \sm{3.31} & \sm{2.01} & \sm{4.71} & \sk{3.81} & \sm{1.28} & \sk{59.25} \\
                \midrule
                \multirow{6}{5em}{Relative response rate} & Content & \sperc{8.96} & \sperc{2.76} & \sperc{0.43} & \sperc{1.27} & \sperc{2.30} & \sperc{5.49} & \sperc{3.76} & \sperc{4.97} & \sperc{4.21} & \sperc{0.13} \\
                 & NSP & \sperc{6.15} & \sperc{0.70} & \sperc{0.59} & \sperc{1.70} & \sperc{5.69} & \sperc{9.33} & \sperc{12.54} & \sperc{8.56} & \sperc{23.46} & \sperc{0.29} \\
                 & Educational & \sperc{1.35} & \sperc{0.83} & \sperc{0.08} & \sperc{1.11} & \sperc{0.42} & \sperc{0.38} & \sperc{0.96} & \sperc{1.47} & \sperc{0.07} & \sperc{0.06} \\
                 & Non-Profit & \sperc{5.48} & \sperc{6.79} & \sperc{0.92} & \sperc{5.12} & \sperc{14.64} & \sperc{9.49} & \sperc{19.99} & \sperc{{0}} & \sperc{6.06} & \sperc{0.36} \\
                 & ISP & \sperc{45.88} & \sperc{13.48} & \sperc{21.18} & \sperc{13.29} & \sperc{36.21} & \sperc{44.28} & \sperc{30.89} & \sperc{11.29} & \sperc{37.07} & \sperc{24.97} \\
                 & Full & \sperc{26.85} & \sperc{8.51} & \sperc{5.58} & \sperc{4.46} & \sperc{18.46} & \sperc{34.31} & \sperc{13.32} & \sperc{7.84} & \sperc{15.46} & \sperc{1.05} \\
                \midrule
                \multirow{6}{5em}{Aliased prefix ratio} & Content & \sperc{1.75} & \sperc{6.14} & \sperc{35.14} & \sperc{0.21} & \sperc{0.38} & \sperc{0.15} & \sperc{0.12} & \sperc{0.37} & \sperc{0.31} & \sperc{40.94} \\
                 & NSP & \sperc{1.29} & \sperc{11.18} & \sperc{34.53} & \sperc{0.14} & \sperc{0.21} & \sperc{0.18} & \sperc{1.43} & \sperc{0.11} & \sperc{1.84} & \sperc{44.21} \\
                 & Educational & \sperc{0.66} & \sperc{14.92} & \sperc{17.81} & \sperc{0.73} & \sperc{4.26} & \sperc{0.44} & \sperc{0.07} & \sperc{0.08} & \sperc{0.63} & \sperc{50.06} \\
                 & Non-Profit & \sperc{1.78} & \sperc{2.22} & \sperc{13.89} & \sperc{2.53} & \sperc{0.02} & \sperc{0.13} & \sperc{21.02} & \sperc{{0}} & \sperc{11.39} & \sperc{41.67} \\
                 & ISP & \sperc{0.68} & \sperc{4.12} & \sperc{31.47} & \sperc{0.28} & \sperc{0.17} & \sperc{0.09} & \sperc{0.04} & \sperc{2.82} & \sperc{0.63} & \sperc{22.13} \\
                 & Full & \sperc{2.26} & \sperc{12.77} & \sperc{42.98} & \sperc{0.30} & \sperc{0.28} & \sperc{0.21} & \sperc{0.08} & \sperc{0.71} & \sperc{0.99} & \sperc{43.30} \\
                \midrule
                \multirow{6}{5em}{Candidate ASes} & Content & \sk{1.03} & \sk{1.22} & \sk{5.14} & \sk{4.03} & \num{969} & \num{884} & \sk{1.08} & \num{675} & \sk{2.64} & \sk{3.13} \\
                 & NSP & \sk{2.59} & \sk{2.98} & \sk{5.21} & \sk{7.16} & \sk{1.87} & \sk{1.74} & \sk{2.00} & \sk{1.34} & \sk{1.31} & \sk{6.32} \\
                 & Educational & \num{819} & \sk{1.33} & \num{980} & \sk{2.36} & \num{493} & \num{473} & \num{627} & \num{572} & \num{469} & \sk{2.50} \\
                 & Non-Profit & \num{430} & \num{377} & \num{215} & \sk{1.50} & \num{263} & \num{264} & \num{323} & \num{0} & \num{147} & \sk{3.29} \\
                 & ISP & \sk{2.10} & \sk{1.77} & \sk{3.86} & \sk{10.08} & \sk{3.26} & \sk{2.82} & \sk{4.25} & \sk{1.47} & \sk{10.02} & \sk{6.06} \\
                 & Full & \sk{4.39} & \sk{4.04} & \sk{6.19} & \sk{20.99} & \sk{10.22} & \sk{10.43} & \sk{16.82} & \sk{3.87} & \sk{19.65} & \sk{7.64} \\
                \midrule
                \multirow{6}{5em}{Responsive ASes} & Content & \num{186} & \num{121} & \sk{1.24} & \sk{1.04} & \num{668} & \num{654} & \num{803} & \num{207} & \sk{1.05} & \num{62} \\
                 & NSP & \num{466} & \num{97} & \sk{1.23} & \sk{2.06} & \sk{1.40} & \sk{1.41} & \sk{1.59} & \num{557} & \num{357} & \num{292} \\
                 & Educational & \num{230} & \num{101} & \num{250} & \num{538} & \num{334} & \num{354} & \num{429} & \num{145} & \num{57} & \num{212} \\
                 & Non-Profit & \num{205} & \num{62} & \num{60} & \num{311} & \num{179} & \num{194} & \num{222} & \num{0} & \num{44} & \num{576} \\
                 & ISP & \num{240} & \num{169} & \num{881} & \sk{3.65} & \sk{2.13} & \sk{1.76} & \sk{3.15} & \num{322} & \sk{3.82} & \num{220} \\
                 & Full & \num{618} & \num{336} & \num{814} & \sk{10.97} & \sk{6.52} & \sk{5.52} & \sk{10.94} & \num{844} & \sk{7.16} & \num{252} \\
                \midrule
                \multirow{6}{5em}{Coverage of seed ASes} & Content & \sperc{17.77} & \sperc{11.56} & \sperc{118.43} & \sperc{99.14} & \sperc{63.80} & \sperc{62.46} & \sperc{76.70} & \sperc{19.77} & \sperc{100.76} & \sperc{5.92} \\
                 & NSP & \sperc{23.57} & \sperc{4.91} & \sperc{62.11} & \sperc{104.05} & \sperc{70.71} & \sperc{71.52} & \sperc{80.58} & \sperc{28.17} & \sperc{18.06} & \sperc{14.77} \\
                 & Educational & \sperc{38.40} & \sperc{16.86} & \sperc{41.74} & \sperc{89.82} & \sperc{55.76} & \sperc{59.10} & \sperc{71.62} & \sperc{24.21} & \sperc{9.52} & \sperc{35.39} \\
                 & Non-Profit & \sperc{65.92} & \sperc{19.94} & \sperc{19.29} & \sperc{100.00} & \sperc{57.56} & \sperc{62.38} & \sperc{71.38} & \sperc{{0}} & \sperc{14.15} & \sperc{185.21} \\
                 & ISP & \sperc{5.66} & \sperc{3.99} & \sperc{20.79} & \sperc{86.10} & \sperc{50.31} & \sperc{41.58} & \sperc{74.28} & \sperc{7.60} & \sperc{90.09} & \sperc{5.19} \\
                 & Full & \sperc{3.63} & \sperc{1.97} & \sperc{4.78} & \sperc{64.48} & \sperc{38.34} & \sperc{32.42} & \sperc{64.31} & \sperc{4.96} & \sperc{42.10} & \sperc{1.48} \\
                \midrule
                \multirow{6}{5em}{Number of newly covered ASes} & Content & \num{27} & \num{8} & \sk{1.14} & \num{268} & \num{20} & \num{14} & \num{23} & \num{0} & \num{296} & \num{44} \\
                 & NSP & \num{52} & \num{32} & \sk{1.05} & \num{531} & \num{66} & \num{49} & \num{25} & \num{0} & \num{70} & \num{255} \\
                 & Educational & \num{32} & \num{15} & \num{231} & \num{128} & \num{12} & \num{24} & \num{15} & \num{0} & \num{10} & \num{191} \\
                 & Non-Profit & \num{65} & \num{7} & \num{57} & \num{109} & \num{15} & \num{18} & \num{10} & \num{0} & \num{13} & \num{555} \\
                 & ISP & \num{18} & \num{6} & \num{688} & \num{639} & \num{26} & \num{13} & \num{10} & \num{0} & \sk{1.18} & \num{161} \\
                 & Full & \num{55} & \num{3} & \num{100} & \num{359} & \num{24} & \num{6} & \num{16} & \num{0} & \num{286} & \num{50} \\
                \bottomrule
            \end{tabularx}
        \end{table*}

\label{lastpage}

\end{document}